# Superconductivity at 5 K in potassium doped phenanthrene

X. F. Wang[1], R. H. Liu[1], Z. Gui[2], Y. L. Xie[1], Y. J. Yan[1], J. J. Ying[1], X. G. Luo[1] and X. H. Chen[1]

1. Hefei National Laboratory for Physical Sciences at Microscale and Department of Physics, University of Science and Technology of China, Hefei, Anhui 230026, P. R. China

2. State Key Laboratory of Fire Science, University of Science and Technology of China, Hefei, Anhui 230026, P. R. China

**Organic materials are believed to be potential superconductor with high transition temperature ($T_C$). Organic superconductors mainly have two families: the quasi-one dimensional (TMTSF)$_2$X and two dimensional (BEDT-TTF)$_2$X (Ref. 1 and 2), in which TMTSF is tetramethyltetraselenafulvalene ($C_{10}H_{12}Se_4$) and BEDT-TTF or "ET" is bis(ethylenedithio)tetrathiafulvalene ($C_{10}H_8S_8$). One key feature of the organic superconductors is that they have π-molecular orbitals, and the π-electron can delocalize throughout the crystal giving rise to metallic conductivity due to a π-orbital overlap between adjacent molecules. The introduction of charge into $C_{60}$ solids and graphites with π-electron networks by doping to realize superconductivity has been extensively reported[3,4]. Very recently, superconductivity in alkali-metal doped picene with π-electron networks was reported[5]. Here we report the discovery of superconductivity in potassium doped Phenanthrene with $T_C \sim 5$ K. $T_C$ increases with increasing pressure, and the pressure of 1 GPa leads to an increase of 20% in $T_C$, suggesting that the potassium doped phenanthrene shows unconventional superconductivity. Both phenanthrene and picene are polycyclic aromatic hydrocarbons, and contain three and five fused benzene rings, respectively. The ribbon of fused benzene rings is part of graphene. Therefore, the discovery of superconductivity in**

**$K_3$Phenanthrene produces a novel broad class of superconductors consisting of fused hydrocarbon benzene rings with π-electron networks. The fact that $T_C$ increases from 5 K for $K_x$Phenanthrene with three benzene rings to 18 K for $K_x$picene with five benzene rings suggests that such organic hydrocarbons with long benzene rings is potential superconductor with high $T_C$.**

Organic superconductors exhibit many interesting phenomena, including low dimensionality, strong electron-electron and electron-phonon interactions and the proximity of antiferromagnetism, insulator states and superconductivity. These organic superconductors consist of open-shell molecular units which are the results of a partial oxidation and reduction of the donor and acceptor molecules in the crystal-growth process. It is the unpaired electron residing in the π-molecular orbital of the donor unit which is responsible for the electronic properties of these charge transfer salts. The π-molecular orbital plays a very important role for superconductivity in $C_{60}$ solids and graphite superconductors, too. One of the key features for organic superconductors is the low dimensionality of the materials. Another important feature is a strong coupling of the charge carriers to the lattices. Organic superconductors provide good system to study the interplay of strong electron-electron and electron-phonon interactions in low-dimensional system. It is striking that the close proximity of superconductivity to a magnetically-ordered state happens in organic superconductors[6,7], similar to the other strongly-correlated electron systems: the heavy fermion metals[8] and high-temperature superconductors including cuprates[9] and pnictides[10,11]. Therefore, it challenges the understanding of superconductivity. It is found that the $T_C$ increases with the expansion of the lattice in alkali-metal doped $C_{60}$. It is expected by the Bardeen, Cooper and Schrieffer (BCS) theorem because the expansion of lattice leads to an enhancement of the density of states (DOS) on the Fermi surface. However, the body-centered-cubic fulleride $Cs_3C_{60}$ is found to be a true Mott Jahn Teller insulator, and it possesses both ingredients required by the strongly correlated theory[12,13]. The insulator $Cs_3C_{60}$ can be turned into superconductor by pressure[14]. It is amazing that the superconductivity induced by pressure in $Cs_3C_{60}$ is related to the antiferromagnetic Mott insulator[12,13], similar to the cuprate. For the graphite superconductors with π-electron networks, $C_6Yb$ and $C_6Ca$ show superconductivity at 6.5 and 11.5 K[15], respectively. Such high transition temperatures are unprecedented and have not been explained by a simple phonon mechanism for the superconductivity. Therefore, it indicates that superconductivity for the superconductors with π-electron networks is not simply explained by BCS theory. Recently, the potassium doped picene ($C_{22}H_{14}$) with fused five benzene rings shows superconductivity at the temperature as high as 18 K. Therefore, it is significant to study the electronic properties of the other phenacenes, the family of aromatic hydrocarbons to which picene belongs. If superconductivity can be induced by introducing carriers into other phenacenes, this raises the prospect of producing a new family of superconductor that has properties as exciting as those their π-electron antecedents. Amazingly, we discover superconductivity in potassium doped phenanthrene ($C_{14}H_{10}$) with three benzene rings.

Phenanthrene is a polycyclic aromatic hydrocarbon and one of the phenacenes, and has a similar structure to that of picene. Only difference between phenanthrene and picene is that they have different fused benzene rings of three and five, respectively. High purity K metal (98%, Sinopharm Chemical Reagent Co.,Ltds) and phenanthrene (98%, Alfa Aesar) were used as starting materials. The K metal was cut into small pieces (1 mm) and then mixed with phenanthrene in the chemical stoichiometry. The mixture was loaded into a quartz tube (Φ 10 mm) later. The quartz tube was sealed under pressure of $10^{-4}$ Pa. The temperature is heated to 200 °C in 40 minutes and was kept at 200°C for 24 hours. At 200 °C, both phenanthrene (m.p. 101 °C) and potassium (m.p.

63.65 °C) are in liquid state to speed up the chemical reaction and insure the homogeneity of the sample. The powder sample shows uniform dark black color, which is totally different from the pure white color of phenanthrene. The appearance of samples before and after annealing was shown as inset of Fig 1(b). $K_x$phenanthrene is quite sensitive to air, and it will be decomposed in air within few minutes. All the processes except for annealing were done in glove box with the oxygen and moisture level less than 1 ppm.

The molecular structure and crystal structure of phenanthrene were shown in inset of Fig 1(a). Both phenanthrene and picene molecules are extended phenanthrene-like structure motif. The phenanthrene contains three fused benzene rings, while picene contains five. The ribbon-like molecular structure can be regarded as part of sheet of graphene. The lattice parameters are determined by X-Ray diffraction. Since $K_x$phenanthrene is extremely sensitive to oxygen and moisture, the powder sample was wrapped in Mylar film (3511 KAPTON, thickness: 8 μm) during X-Ray diffraction measurement. The wrapped film containing samples was placed at Micro X-Cell (Number 3529 , SPEXCertiPrep Ltd). The X-Cell setup was fixed to the sample shelf of X-ray apparatus (TTR-III theta/theta rotating anode X-ray Diffractometer, Japan) by plasticine. X-ray diffraction pattern was obtained in the 2-theta range of 5°-65° with a scanning rate of 1° per minute. The lattice constants were determined by UnitCell software. Fig 1(a) is the X-Ray diffraction pattern of phenanthrene. As shown in top panel, the phenanthrene crystallizes in *P2₁* symmetry with lattice parameters: a=8.453 Å, b=6.175Å, c=9.477 Å, β=98.28°. All peaks in XRD pattern can be well indexed with the crystal structure mentioned above. Miller indices are marked in the pattern. The lattice constants obtained here are consistent with the results reported before[16]. The crystal comprises layers stacked in *c* direction, where each layer (in a-b plane) contains phenanthrene molecules arranged in a herringbone structure, as shown in the inset of Fig 1(a). Fig

1(b) shows the X-Ray diffraction pattern of superconducting $K_x$phenanthrene. The hump near 15º and obvious background signal are due to Mylar Film. The impurity phases KH and KOH·$H_2$O were identified in the XRD pattern. The KH impurity phase arises from that some excessive potassium reacts with amount active hydrogen of phenanthrene molecule. The KOH·$H_2$O could be induced by very small leak of air into the sealed Mylar film during measurement of XRD. Those K intercalated samples are very sensitive to air. When they were exposed in the air, the recovered phenanthrene was observed in XRD pattern due to K de-intercalating from $K_x$phenanthrene (see the supplement information). This phenomenon is quite similar to alkali metal intercalated graphite, where the intercalated alkali metals are easily de-intercalated from the graphite interlayer when they are exposed in air[17-19]. As potassium intercalated into the phenanthrene, the lattice constants change to a=8.650 Å, b=5.963 Å, c=9.297Å, β=100.16° and the unit cell volume also contracts to 472.0 Å$^3$ from 489.6 Å$^3$. The evolution of lattice constants with K doping is quite similar to the case of $K_x$picene reported by Mitsuhashi et al[6]. The remarkable feature is that the lattice constant *c* and *b* shrink, while the *a* expands with intercalating K. Contrasting to alkali atoms intercalated pentacene[20], where the lattice constant of *c* axis is significantly expanded due to alkali metal atoms intercalating between the molecular layers of pentacene. Here the intercalation of K atoms should be intralayer intercalation, so that it leads the expansion of the *a* and shrink of the *c* and *b*, similar to that of $K_x$picene[6]. Those intralayer intercalations are also quite different from (BEDT-TTF)$_2$I$_3$[21] (BEDT-TTF=bis(ethylenedithio) tetrathiafulvalene) and (TMTSF)$_2$PF$_6$[22] (TMTSF=Tetramethyltetraselenafulvalene) superconductors. In superconducting (BEDT-TTF)$_2$I$_3$ and (TMTSF)$_2$PF$_6$, the electron donor and acceptor layers form quasi two dimensional layers which is layer-by-layer stacked.

The magnetization measurement was performed on SQUID MPMS (Quantum Design). The sample was placed into a Teflon cell (Quantum Design), then the cell was sealed by coal oil in glove box. Figure 2(a) shows the magnetic susceptibility $\chi$ as a function of temperature for the powder sample $K_x$phenanthrene in the zero-field cooling (ZFC) and field cooling (FC) measurement procedures under a magnetic field (H) of 10 Oe. The magnetic susceptibility $\chi$ versus T plot shows a sharp decrease below 4.7 K in ZFC measurements, and the temperature corresponding to the sharp decrease is defined as the superconducting transition temperature ($T_C$). The diamagnetic $\chi$ in the ZFC and FC measurements can be assigned to the shielding and Meissner effects, respectively. As seen from Fig. 2(**a**), $T_C$ was determined to be 4.7 K for the powder sample-C. The shielding fraction is about 4.3% for the powder sample-C. It has to be mentioned that the superconductivity disappears immediately when the sample is exposed to air. As discussed by Mitsuhashi et al[5], the small shielding fraction in powder sample could arise from the penetration of magnetic field into superconducting phase because of smaller size of crystallites than the London penetration depth. The superconducting $K_x$phenanthrene powder sample-C was pressed into a pellet, and the shielding fraction is increased to about 11.6% for the pellet sample. The shielding fraction can be increased when the powder sample was pressed to form a pellet, suggesting that the superconductivity arises from bulk superconductivity. One can see that the superconducting transition is very sharp and transition width is less than 0.6 K. The shielding fraction of the $K_x$phenanthrene is larger than that of $K_x$picene[5]. The lower critical magnetic field $H_{C1}$~150 Oe was estimated from the M versus H plot for sample-C at 2 K, less than $H_{C1}$~380 Oe in $K_{3.3}$picene with $T_C$~18 K at 5 K[5]. The measurements of magnetization versus field (inset of Fig.2a) indicate the magnetic hysteresis expected in a bulk superconductor with flux pinning. It suggests that $K_x$phenanthrene should be a type-II superconductor. Figure 2(b) shows temperature dependence of $\chi$ under different H for sample-C in the ZFC measurement. One can see that the diamagnetic signal disappears distinctly with increasing H, but there was an obvious drop of $\chi$ at

4.15 K even at 1000 Oe. Magnetic field dependence of $T_C$ was obtained based on the data shown in Fig 2(b), and the plot of upper critical field $H_{c2}$ versus T is shown in the inset of Fig.2b.

Figure 3(a) presents the temperature dependent magnetization for the powder sample $K_x$phenanthrene (sample-A) under the pressures ranging from ambient pressure to 1 GPa. The most notable result is that both $T_C$ and shielding fraction increase rapidly with increasing the pressure from ambient pressure to 1GPa. $T_C$ is enhanced from 4.7 K at ambient pressure to 5.9 K at 1 GPa. The pressure dependence of $T_C$ for sample-A is shown in Figure 3(b). The positive $d(T_C/T_C(0))/dP$ ~$0.26$ $GPa^{-1}$ at low pressure range (less than 1 GPa) in $K_x$phenanthrene is similar to graphite intercalation compounds (GICs)[23]. Such positive dependence would not be expected within a conventional weak coupling phonon model. It should be noted that the magnetic susceptibilities in normal states for $K_x$phenanthrene are considerably large, and non-superconducting $K_x$phenanthrene shows the Curie-type behavior. Similar behavior has been observed in $K_x$picene. It implies that there exist local spins in these materials. It is well known that a positive pressure effect on superconductivity is widely observed in $Cs_3C_{60}$[12], high-$T_C$ cuprate compounds[24] and iron-based superconductors[25] as a consequence of the suppression of antiferromagnetism (AFM). Such a positive pressure effect on $T_C$ in aromatic hydrocarbon superconductors may be due to the suppression of the localized spin interaction by pressure.

We prepared the series of the samples $K_x$phenanthrene with different potassium content. It is found that only the sample with nominal composition $K_3$phenanthrene shows superconductivity, and all other samples with x deviation from 3 (even slightly deviating from 3, for example 2.9 or 3.1) do not show superconductivity (see supplementary information). It seems that the superconducting phase for $K_x$phenanthrene superconductor is a line phase $K_3$phenanthrene. This is totally different from the case of potassium doped picene in which the samples $K_x$picene with $2.6 \leq x \leq 3.3$ show superconductivity and there exist two superconducting phases with $T_C$=7 and 18 K, respectively. There exists a big family of ploycyclic aromatic compounds with an extended phenanthrene-like structural motif designated as [$n$]phnenacens, where $n$ is the number of fused benzene rings. [$n$]phenacene molecules are related to layers of graphene in the way that ribbons are related to sheets. Therefore, discovery of superconductivity in $K_x$phenanthrene ($K_xC_{14}H_{10}$) besides the superconductor $K_x$picene opens a new broad family of superconductors that consists of aromatic hydrocarbons. To search for superconductors in such family of [$n$]phenacene molecules and to study their physical properties are very important to understand their superconductivity. High quality superconducting sample, even single crystal form is necessary for determining of structure and the exact superconducting phase, and for studying their physical and chemical properties to reveal the mechanism of superconductivity.

**Methods:**

The magnetization measurement was made by SQUID MPMS (Quantum Design). The sample was placed into Polypropylene powder holder (Quantum Design). Structure identification was carried out by TTR-III theta/theta rotating anode X-ray Diffractometer (Japan). The sample was packaged by Mylar film (SPEX CertiPrep Group) during X-Ray diffraction measurement. The samples should be protected from air or moisture carefully. When the sample meets oxygen or moisture, the superconductivity disappears immediately. The magnetization under pressure was measured by incorporating a copper-beryllium pressure cell (EasyLab, UK) into SQUID MPMS (Quantum Design). The sample was firstly placed in a teflon cell with coal oil (EasyLab, UK) as the pressure media. Then the teflon cell (EasyLab, UK) was set in copper-beryllium pressure cell for magnetization measurement. The contribution of background magnetization against the sample at 2 K was less than 5%.

**Acknowledgements:** This work was supported by the Natural Science Foundation of China, and by the Ministry of Science and Technology of China and Chinese Academy of Sciences.

**Author Contributions:** X.H.C. designed and coordinated the whole experiment, analyzed the data and wrote the paper. X.F.W. contributed to the synthesis, and XRD and magnetization measurements. R.H.L. did the structure analysis and analyzed the data. Z.G., Y.L.X, Y.J.Y, J.J.Y, X.G.L. did some of magnetization and high pressure measurements.

**Author information:** Correspondence and requests for materials should be addressed to X.H.C. (chenxh@ustc.edu.cn).

**Figure Captions:**

**Figure 1 X-Ray diffraction pattern for phenanthrene and $K_x$phenanthrene. (a)**: XRD pattern for the phenanthrene purchased from Alfa Aesar. The molecular structure and crystal structure of phenanthrene are shown in the inset. **(b)**: XRD pattern for the superconducting $K_3$phenanthrene sample. The hump near 15º is caused by Mylar film. The impurity phase KH (potassium hydride) and KOH·$H_2$O are observed in the XRD pattern. The inset shows the appearance for the pristine phenanthrene (white) and $K_x$phenanthrene (black).

**Figure 2 Temperature dependence of magnetization χ for $K_x$phenanthrene. (a)**: χ versus T plots for $K_x$phenanthrene with $T_C$=4.7K for the powder sample-C (main) and the pellet sample by pressing $K_x$phenanthrene powder (right inset). The shielding fraction is 4.3% for the powder sample-C and 11.6% for the pellet sample. M versus H plots for sample-C at 2K in the left inset. **(b)**: χ versus T plots for the powder sample-C in the zero-field cooling (ZFC) measurements under different H. The H versus $T_C$ plot is shown in the inset. The onset temperature ($T_C^{onset}$) is defined as $T_C$.

**Figure 3 Pressure dependence of the superconducting transition $T_C$. (a):** Magnetization χ as a function of temperature for the powder sample-A under the pressures of p=0, 0.2, 0.4, 0.6, 0.8, 1.0 GPa in ZFC measurements. **(b):** Evolution of $T_C$ with pressure for the superconducting $K_x$phenanthrene.

# Figure 1

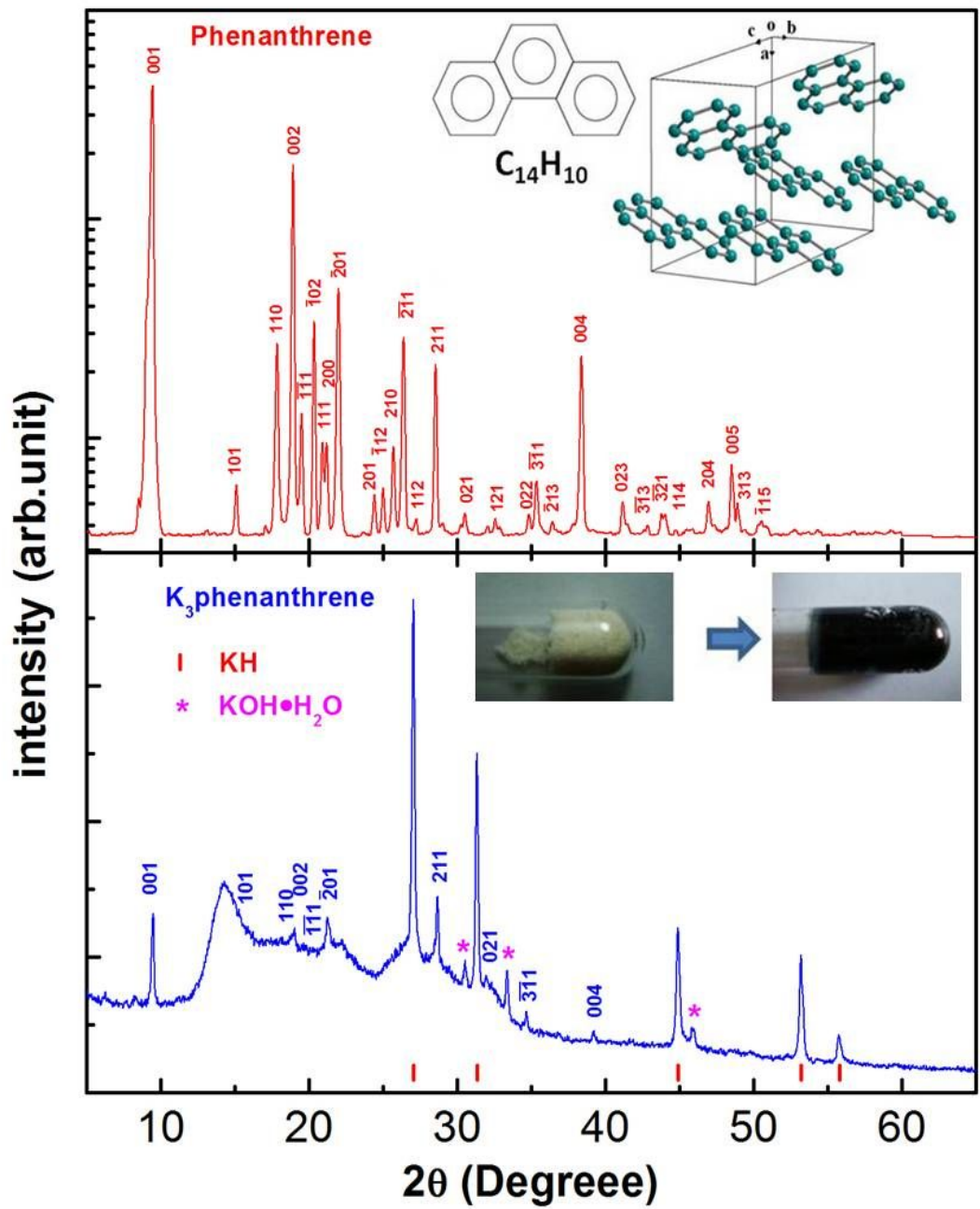

Figure 2

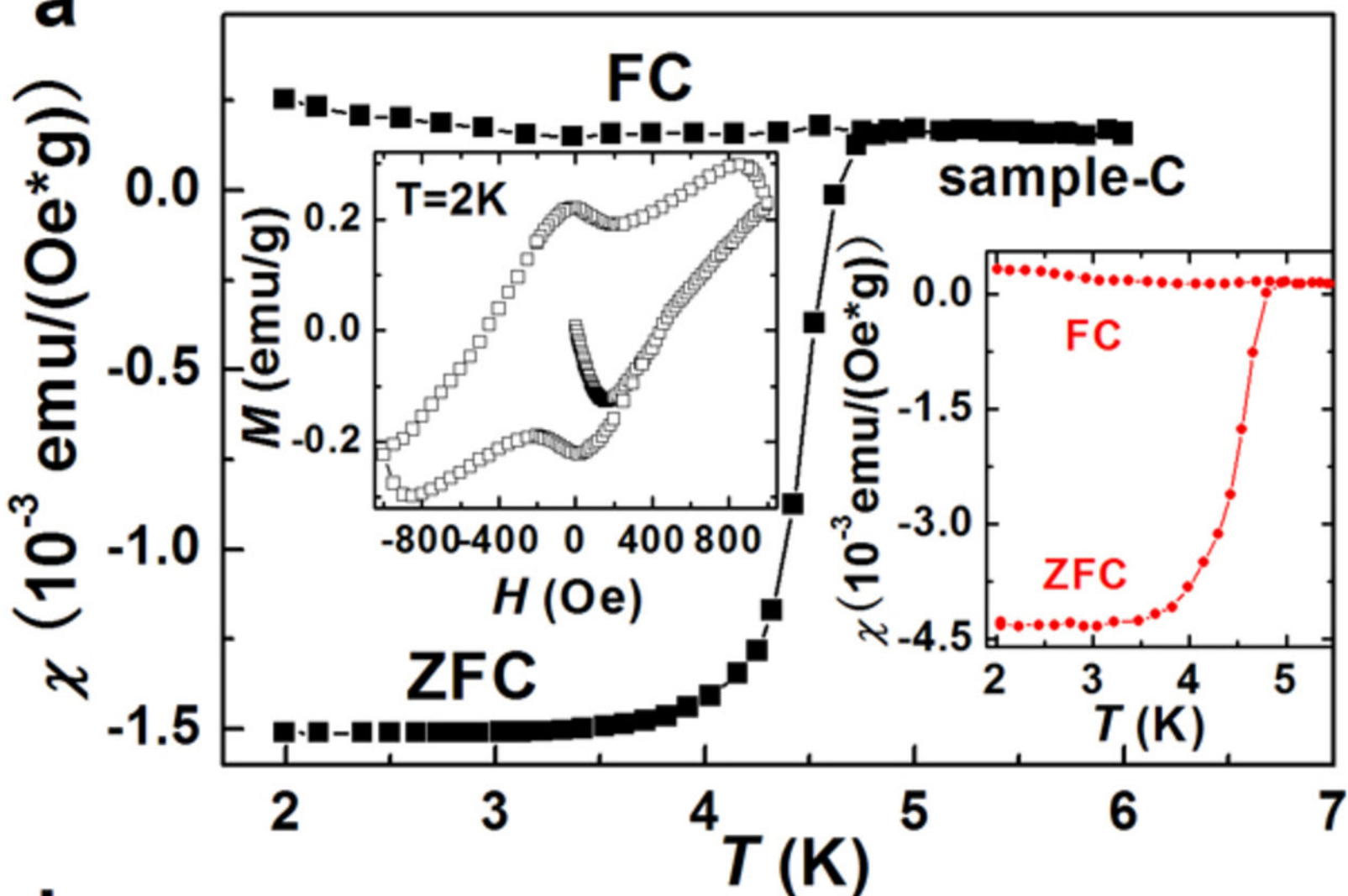

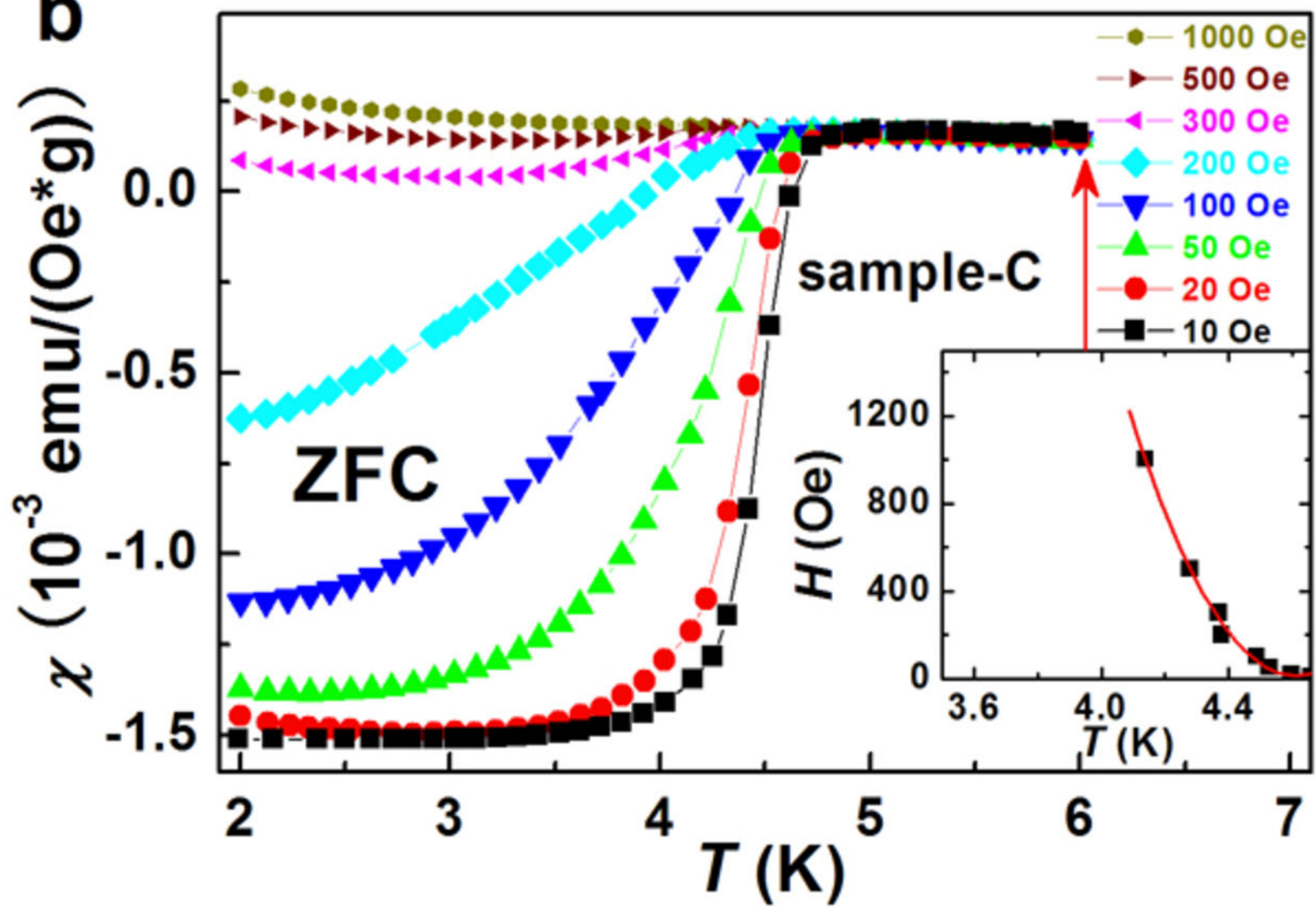

Figure 3

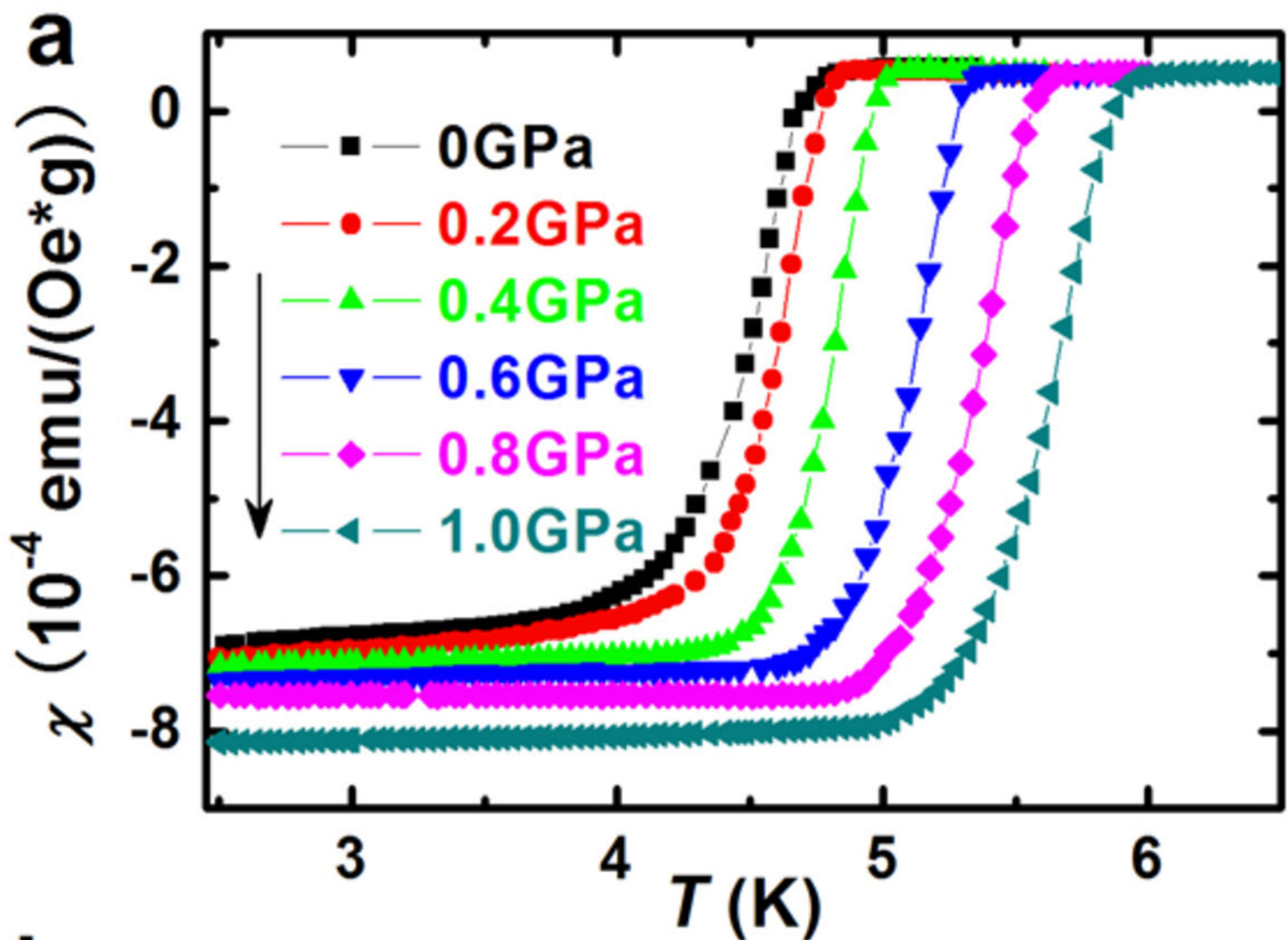

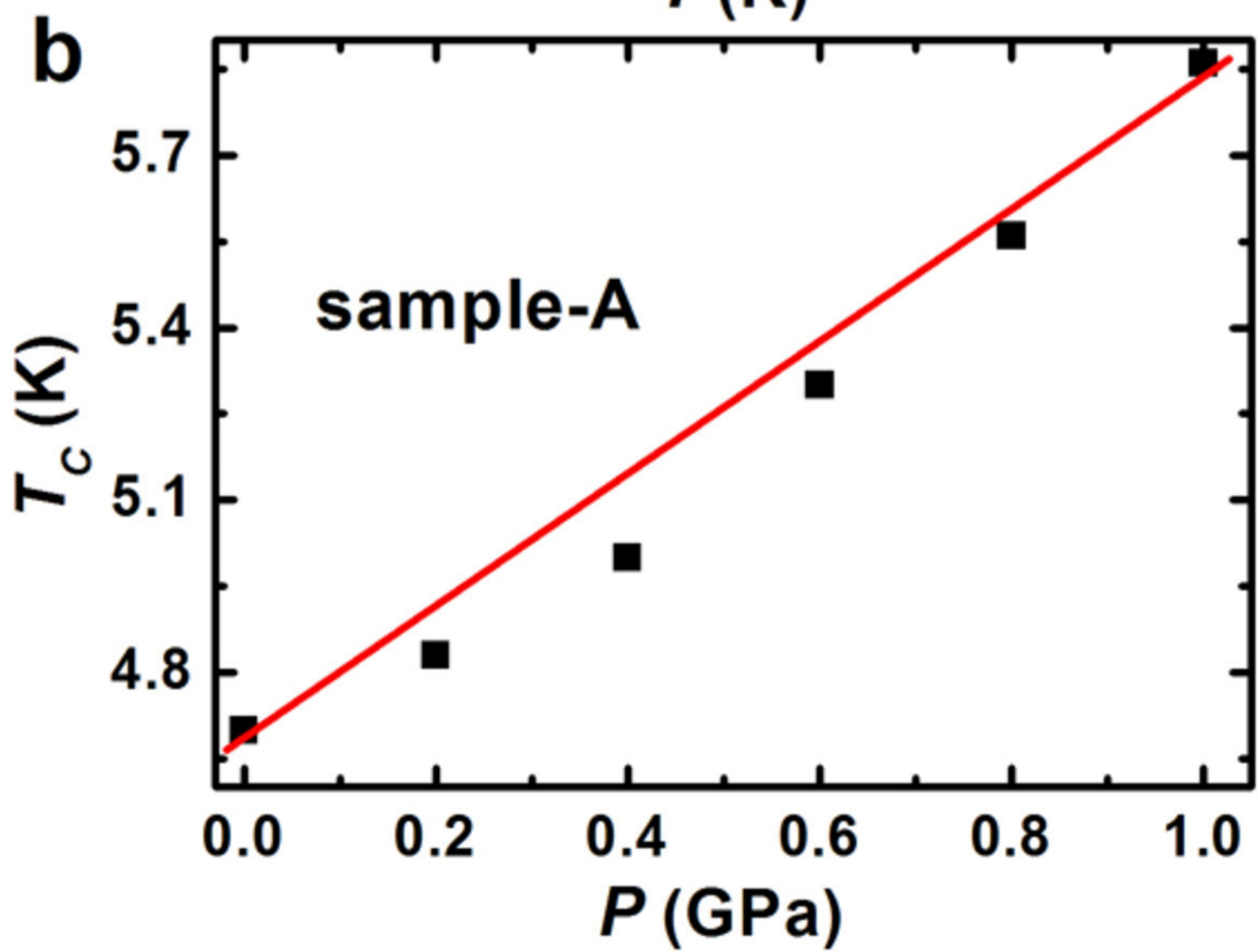

## Supplementary Information:

**The structure**

**Figure S1** shows the XRD patterns for the three different batches of superconducting samples with the same reaction condition, which are wrapped in Mylar film during X-Ray diffraction measurement. In XRD patterns, in addition to the phase $K_x$phenantrene, there exists KH and a little $KOH \cdot H_2O$ impurity, whose Bragg reflections are indicated by the vertical short lines. The $KOH \cdot H_2O$ could be induced by the slight leak of air into the sealed Mylar film. The superconducting phase should be $K_x$phenantrene because the impurity phases of KH and $KOH \cdot H_2O$ are non-superconducting. As we removed the Mylar film and exposed the powder samples to air, the black powder products changed to white quickly. After exposing the powder to air for 10 minutes, we performed XRD test again and obtained XRD patterns as shown in **Fig. S2**. **Figure S2** indicates that the intercalated K in $K_x$phenantrene is released in air and the main phase of phenanthrene is observed in XRD pattern after K de-intercalating from $K_x$phenantrene. In addition, one peak of organic material $C_{16}H_{16}O_2$ is observed. At the same time, the most of KH reacts with moist air in short time to form the $KOH \cdot H_2O$. The potassium de-intercalating from $K_x$phenantrene in air is consistent with the fact that the $K_x$phenantrene loses the superconductivity when it touches a little air. Such instability of $K_x$phenantrene in air is quite similar to that of alkali-metal-intercalated graphite[1-3].

**Superconductivity for different batches of $K_x$phenanthrene:**

We prepared nine batches of $K_x$phenanthrene in the same condition. It is found that only the sample with nominal composition $K_3$phenanthrene is superconducting. The XRD patterns of superconducting samples were shown in **Fig. S1**. The parameters of superconductivity for those $K_x$phenanthrene is displayed in **Table S1**, where $T_C$ and shielding fractions were obtained from the magnetic susceptibility measurements. Although the superconducting fraction is quite different from batch to batch among the three superconducting samples, the superconducting critical temperature $T_C$ changes less than 5%. The difference of the superconducting fraction may arise from the different size of the crystallite. The nominal composition is deviated from $K_3$phenanthrene (for example: $K_{2.9}$phenanthrene, $K_{3.1}$phenanthrene etc), the superconductivity completely disappears. It seems that the superconducting phase is a line phase, which is different from $K_x$picene $(2.6 \leq x \leq 3.3)^4$.

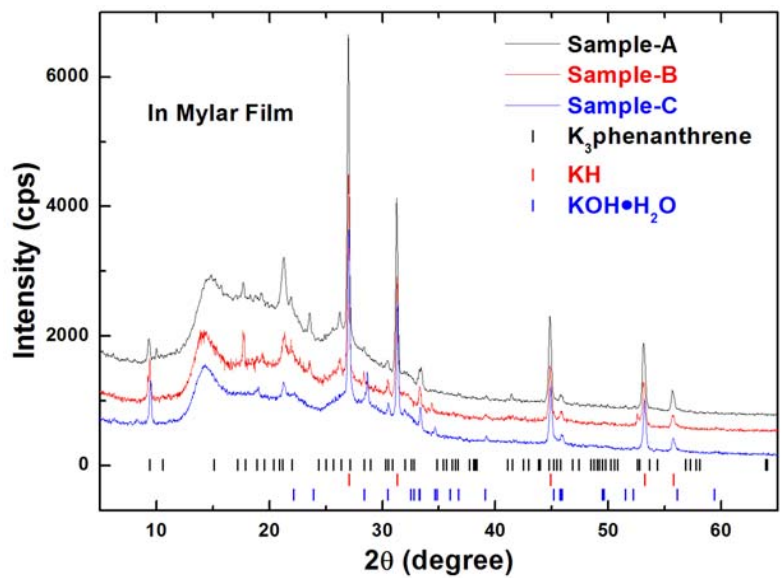

**Figure S1**: The X-ray diffraction patterns for three different batches of superconducting samples with the same reaction condition. The vertical short lines indicate the Bragg reflections of the possible phases existing in the reaction products. The main phases are $K_x$phenantrene and KH, respectively. A little KOH · $H_2O$ may come from KH reacting slightly with moist air when XRD data was collected in Mylar film. The hump near 15° and obvious background signal arises from Mylar Film.

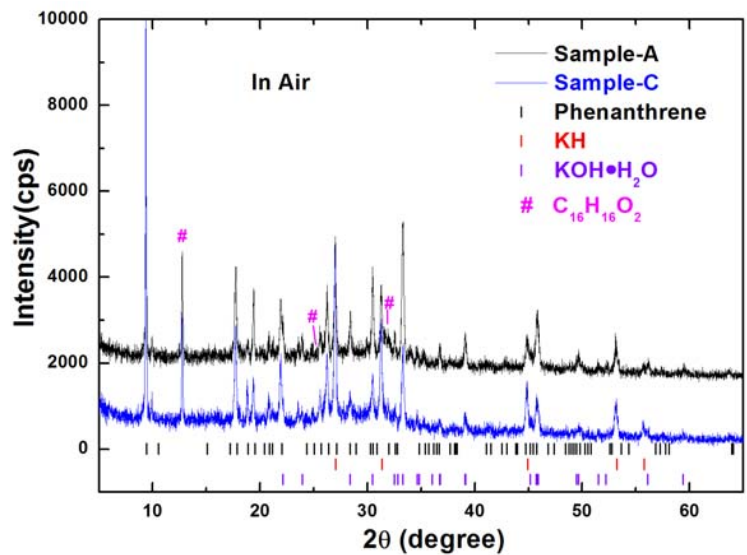

**Figure S2** : The XRD patterns for the two samples after being exposed in air for 10 minutes. Main phase is phenanthrene, which is recovered from K$_x$phenanthrene after putting the products in air. The different batches of powders show the almost same XRD patterns after being exposed to air. The XRD reflection positions of phenanthrene and the impurity phases KOH·H$_2$O, KH and the possible C$_{16}$H$_{16}$O$_2$ are shown using by the vertical short lines.

Table S1. List of nine batches of $K_x$Phenenthrene samples.

| N.O. | Nominal composition | Annealing temperature | Annealing time | Onset SC transition | Shielding fraction |
|---|---|---|---|---|---|
| Sample A | $K_3$Ph | 200°C | 20 hours | 4.7 K | 1.9% |
| Sample B | $K_3$Ph | 200°C | 20 hours | 4.9 K | 1.1% |
| Sample C | $K_3$Ph | 200°C | 20 hours | 4.7 K | 4.3% |
| Sample D | $K_{2.5}$Ph | 200°C | 20 hours | NO SC | - |
| Sample E | $K_{2.8}$Ph | 200°C | 20 hours | NO SC | - |
| Sample F | $K_{2.9}$Ph | 200°C | 20 hours | NO SC | - |
| Sample G | $K_{3.1}$Ph | 200°C | 20 hours | NO SC | - |
| Sample H | $K_{3.2}$Ph | 200°C | 20 hours | NO SC | - |
| Sample I | $K_{3.5}$Ph | 200°C | 20 hours | NO SC | - |